\documentclass[showpacs,twocolumn,prb]{revtex4}
\usepackage{amsthm,amsfonts,graphicx,verbatim}
\begin{document}

\title{Transport in Coherent Quantum Hall Bilayers}

\author{D. A. Pesin, and A. H. MacDonald}
\affiliation{Department of Physics, University of Texas at Austin,  Austin TX 78712 USA}

\begin{abstract}

We discuss two phenomenological descriptions of low-current transport in bilayer quantum Hall system with exciton condensates, one based on a Landauer-Buttiker description of Andreev scattering at contacts to coherent  bilayers and one based on a simplified single-parameter {\em p-ology}
description of the weak to strong interlayer coupling crossover.  The Andreev scattering phenomenology in intended to apply when the condensate is well developed and is used to predict current-voltage relationships for a variety of two contact geometries.  We also apply this formalism
to circumstances in which the tunnel current exceeds its critical value and the condensate is time-dependent.
The {\em p-ology} approach is used to establish the universal development of large longitudinal drags, even in homogenous
coherent samples, as the condensate weakens and the Hall drag is reduced.

\end{abstract}
\date{\today}
\maketitle
%\tableofcontents

\section{Introduction}

The ground state of a weak-disorder quantum Hall bilayer has an exciton condensate when the total
Landau level filling factor $\nu_{tot} \sim 1$ and the semiconductor quantum wells are close together.
The condensate state is characterized by spontaneous inter-layer phase coherence.\cite{Fertig1989, GirvinMacDonaldBook}
Condensate properties are often probed most effectively by transport
experiments, both because the two-dimensional electron system is buried beneath the surface of a three-dimensional host and
because these measurements probe\cite{su_to_2008} condensate superfluidity.
Indeed, the presence of a condensate has been inferred on the basis of wide variety of
transport anomalies.\cite{SpielmanPRL2000,KelloggPRL2002, Kellogg2003, KelloggPRL2004, Eisenstein2004, TutucPRL2004, tutuc_giant_2009, Tutuc2005intANDdis, TiemannPRB2008, Tiemann2009Dominant, Finck2008AreaDepCond, Champagne2008FiniteT, Wiersma2004, Yoon2010}
Although these experiments do not directly measure the condensate order parameter, it is
generally acknowledged that they do not allow plausible alternate explanations.  On the other
hand the transport properties of systems in which the condensate is believed to have formed
have largely defied attempts to achieve fully quantitative interpretations.  This conundrum is related
in part to the importance of
edge states in quantum Hall transport\cite{HalperinEdge,Buttiker1988}
and, as we explain in this paper, to the sensitivity of transport measurements to disorder and to
contact properties that are not normally characterized.

The description that
we develop for transport in the deeply coherent regime is based on a microscopic mean-field theory of the
exciton condensate state, on a scattering formulation of transport theory,\cite{Landauer1970,ButtikerPRL1986}
and on a two-channel contact model which accounts explicitly
for the two contact layers.  We show that transport properties can be non-trivial even when the
current is small enough that the coherence order parameter is time-independent.\cite{WenZee,Wen_ZeePRB, RossiPRL2005,SuPRB2010,eastham_critical_2010}
A simplified version of this picture with a single parameter, which we refer to as {\em p-ology}, allows
for virtual occupation of a pair-breaking edge channel and
can be used to provide an approximate description of the crossover between strong and weak interlayer coupling regimes.

Our paper is organized as follows.  In Section II we explain the general philosophy of our approach, which has
similarities to theories of the non-equilibrium properties of both superconductors~\cite{Kopnin_book}
and magnetic conductors.\cite{Bauer_RMP}
The edge state physics of quantum Hall systems is essential to the way in which the theory is developed.
Indeed, the presence of edge states at the Fermi level, even when
the bulk is gapped, is responsible for many of the complications
which make the transport anomalies difficult to describe quantitatively.
In Section III we detail the Landauer-Buttiker theory that we use to describe transport of the
mean-field-theory quasiparticles, and illustrate it by evaluating two probe conductances for
systems with well developed coherence.
In Section IV we discuss transport in the case of a  non stationary condensate,
focusing on the simplest case in which both coherent and incoherent inter-layer tunneling are absent and
illustrating our picture by calculating the interlayer voltage for a drag experiment with a grounded drag layer.
In Section V we develop
{\em p-ology}, which is intended to describe the influence of
weaker coherence but assumes that quasiparticle transport is still edge dominated.
This phenomenology makes it clear that large longitudinal drags are always present when the
Hall drag voltage drops below its quantized value, as seen in experiment.\cite{KelloggPRL2002,Kellogg2003,tutuc_giant_2009}
Sec. VI contains a summary of our main results and discusses
their relationship to other ideas for explaining the transport properties of coherent bilayers.

\section{Theoretical picture}
In this paper we adopt a Hartree-Fock-like mean-field description of the exciton condensate which is similar to the
mean-field-theories used, generally successfully, to describe transport in metallic ferromagnets and superconductors.
This type of theory has an $O(3)$ vector order parameter constructed in our case from the pseudospin-$1/2$ layer degree of freedom,
and an associated pseudospin effective magnetic field which arises from microscopic exchange interactions.
For a typical bilayer with weak single-particle tunneling, the exchange field exceeds the
single-particle symmetric-antisymmetric level splitting, $\Delta_{SAS}$, by many orders of magnitude.
$\Delta_{SAS}$ is therefore normally negligible in the description of the quasiparticle properties.
The dynamics of the order parameter, on the other hand, is described by a Landau-Lifshitz-like equation which
depends only on an external field term proportional to $\Delta_{SAS}$, on other terms proportional to order parameter derivatives
and on transport currents.\cite{Yang_1996,Nunez2006}  (Characteristic length scales then emerge from a comparison of $\Delta_{SAS}$ and the coefficients of
the order-parameter derivative terms.)
The $\Delta_{SAS}$ term must be present\cite{WenZee, Wen_ZeePRB , RossiPRL2005,SuPRB2010,eastham_critical_2010}
to allow consistently for the net interlayer charge transfer
which occurs in many transport experiments.

Typical electrical measurements performed on a coherent quantum Hall bilayer involve establishing a
source to drain voltage bias which injects transport quasiparticles, and measuring local electrochemical potentials at
positions along the sample perimeter.
When the quantum Hall effect is well developed the transport quasiparticles are localized at the sample edges.
Electrochemical potential measurements therefore often reflect properties of the bulk condensate somewhat indirectly.
Contacts to the coherent bilayer resemble normal metal-superconductor junctions,
at which normal currents are converted into supercurrents.  This conversion between
normal current and supercurrent is accomplished via Andreev scattering.
A similar conversion process occurs in a coherent bilayer;
even for vanishingly small single-particle tunneling a quasiparticle that has a definite layer quantum number in the leads becomes coherently mixed between the layers far away from the leads. The order parameter exchange field provides a broken symmetry contribution to the quasiparticle Hamiltonian with a large inter-layer tunneling amplitude.
This process does not violate charge conservation since an excitonic supercurrent is always launched to compensate\cite{su_to_2008}
for charge transfer between layers.  When the condensate is described collectively,
the full system of equations consists of a Landau-Lifshitz-like equation for the condensate
with a {\em pseudospin-transfer} torque\cite{SuPRB2010,RossiPRL2005} term to account for normal current to exciton super current conversion,
combined with a transport theory for quasiparticles influenced by the condensate.  The two sets of equations need to be solved self-consistently.
At low temperatures, inelastic scattering is weak and quasiparticle transport can be described using the Landauer-Buttiker formalism.
Application of this picture is complicated in
coherent quantum Hall bilayer systems by the role of chiral edge states,
which can carry nearly dissipationless currents by a mechanism completely different from
superfluidity.

%\textbf{Pseudospin transfer torque analogy}

\section{Landauer-Buttiker description of separately contacted bilayers\label{sec:LB}}

Two signature properties of coherent bilayers are i) the quantization of Hall drag resistance at $h/e^2$
(the Hall resistance is quantized whether measured in the layer in which current is flowing or in the
layer in which no current flows), and ii) the absence of longitudinal drag at filling factor
$\nu=1/2$ per layer.  Together the two properties suggest that a quantum Hall effect is established at a
filling factor related to the combined densities of the two layers, and that the quasiparticle current flows
coherently through both layers; there is no layer dependence in chemical potential measurements.
Because of the quantum Hall effect, and its associated bulk gap,
the bilayer coherent quasiparticles must flow at the sample edges.
Since these properties are normally measured in samples in which
$\Delta_{SAS}$ is extremely small, the quasiparticle current in the drag layer
must be exactly compensated by a condensate counterflow supercurrent.

We will be interested in analyzing transport both in four-probe geometries, which include voltage-probe contacts in addition to
the current source and drain contacts, and in two-probe geometries in which the additional voltage probe contacts are
absent.  As we explain, some interesting information about the system can be obtained from two-probe measurements.
 %Quite often in the single layer $\nu=1$ QHE values of two-probe conductance close to $\frac{e^2}{h}$ reported, for leads close to ``ideal''~\cite{Buttiker1988}.
One interesting issue that arises in the two-probe case concerns a theory of the maximum conductance possible when source and drain contacts are connected to opposite layers. Our analysis relates two-probe conductances to the influence of the condensate on electron scattering near the contacts, information which will prove helpful in analyzing four-probe experiments.

An appropriate microscopic model for the {\em normal} contact region in which interlayer coherence is absent
plays an essential role in our analysis.  We assume that the active area of the bilayer in which coherence is established is
incompressible and surrounded by a single current-carrying edge state channel shared between the two layers.\cite{Edge_reconstr}

As a convenience we imagine the leads as consisting of narrow Hall bars with separate $\nu=1$ states
in the two layers and associated ingoing and outgoing states which can be viewed as being located to
some degree on opposite edges of these leads.
We recognize that this is a somewhat idealized model of the incoherent bilayer contacts, but
believe that it is general enough to capture all
the essential physics.  The contacts are indeed realistically viewed as narrow Hall bars, but the
filling factors in each layer depends on the device layout, especially on the arrangement of the main top and
bottom gates.\cite{JPEprivate} Because there is a single channel in the active region, however, we can in principle
always identify a linear combination of lead channels into which electron waves incident from the
sample are transmitted and also a linear combination of lead channels which is responsible
for all transmission.  The incoming and outgoing channels in the leads in our
model should be understood as represented these sample-dependent channels.
With this understanding, even if other channels are present they do not contribute to transport. When the temperature $T \to 0$, this description should apply on the bilayer $\nu_{tot}=1$ Hall plateau, which can in principle extend over the filling factor range $\nu_{tot} \in (0.5,1.5)$.
At finite temperatures there will always be corrections due to thermally excited mobile bulk carriers, and these corrections will be small over a narrower filling factor range.

\begin{figure}
\begin{center}
\includegraphics[scale=0.4,bb=1 0 548 230]{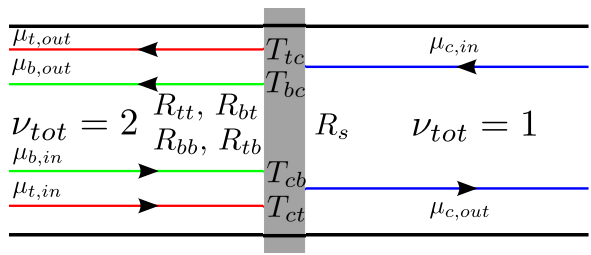}
\caption{(color online) Minimal model of a junction between a normal bilayer region and a $\nu=1$ coherent quantum Hall bilayer.
The normal region has two layers, each with a left-going and right-going channel that can transmit
electrons toward or away from the coherent layer.  On a $\nu=1$ plateau the coherent region
supports a single edge channel with an inter-layer phase relationship that is dictated by the local condensate phase. Shaded regions schematically represent places where scattering in the contacts takes place.
Like a ferromagnet, and unlike a $T=0$ superconductor, the ordered region has conducting quasiparticles.
Unlike typical ferromagnets, the conducting pseudospins are localized at the sample edge and share the same pseudospin state, in the
jargon of magnetism the system is half-metallic.}
\label{fig:edgestates}
\end{center}
\end{figure}

The single edge state in the active area is a certain coherent superposition of the top and bottom edge states, to be specified more completely
below.  Under most circumstances the quasiparticles have equal weight in the two layers
corresponding to the pseudospin order parameter lying in the $XY$-plane.
The phase relationship between quasiparticle components in the two layers is dictated by the local direction of the pseudospin effective Zeeman field.
Occupied edge states therefore add to the condensate strength.
This picture is summarized schematically in~Fig.\ref{fig:edgestates}.
Occupation of the orthogonal quasiparticle state with the opposite interlayer phase would be
pair breaking, {\em i.e.} it would weaken the condensate.
This state has higher energy and does not
participate in low temperature, low bias-voltage transport.

Later we consider the possibility of virtual occupation of the
pair-breaking edge-state channel when the condensate is weak.
Disorder gives rise to a random potential difference
between edges which mixes edge state channels.
This can give rise to memory of the
layer from which an electron has been injected that persists along the edge.
When layer memory is important for the electrical measurements we
need to employ a
different formulation of the scattering transport theory
which has a larger number of free parameters.

The description of scattering at a junction separating two incompressible states, one
representing the contact arms taken to have $\nu_{tot}=2$, and one representing the active area having $\nu_{tot}=1$,
represents a new and interesting example of the application of Landauer-Buttiker transport theory
to the QHE\cite{Buttiker1988} regime.
In this case the two arms can be alternatively thought of as a combined system with two edge states, or
as two separate leads attached to the bulk.
In coherent bilayer transport experiments current is often injected into or drained from only one layer.
We model separate-contacting to a single layer simply by setting the current to zero for one of the layers.
In this way one arm of the contact acts like a voltage probe and the other like a current probe.
In experiment the voltage on the open layer can normally not be read externally, although our
theory does predict its value.
The bulk of the bilayer contains a single edge state, as is appropriate for $\nu_{tot}=1$.
The scattering problem near a contact, depicted in Fig.~\ref{fig:edgestates}, is then defined via transmission probabilities from the upper ``coherent'' edge state to top and bottom edge states in the arms, $T_{tc},\,T_{bc}$, and from top and bottom edge states into the lower coherent edge state, $T_{ct},\,T_{cb}$,  as well as the layer-diagonal reflection probabilities ($R_{tt},\,R_{bb}$), the cross-layer (Andreev) reflection probabilities ($R_{bt},\,R_{tb}$), and finally the probability for a coherent edge state to bypass a lead, $R_s$.
Note that we adopt the convention that the first index of a transmission or reflection probability
denotes the outgoing scattering state channel, while the second index specifies the incoming channel.

Using the above scattering probabilities, one can relate the electrochemical potentials of incoming and outgoing states for the problem illustrated in Fig.~\ref{fig:edgestates}:
\begin{equation}\label{eq:chempots}
  \left(\begin{array}{c}
    \mu_{c,out}\\
    \mu_{t,out}\\
    \mu_{b,out}
  \end{array}\right)=
  \left(\begin{array}{ccc}
    R_s&T_{ct}&T_{cb}\\
    T_{tc}&R_{tt}&R_{tb}\\
    T_{bc}&R_{bt}&R_{bb}
  \end{array}\right)
  \left(\begin{array}{c}
    \mu_{c,in}\\
    \mu_{t,in}\\
    \mu_{b,in}
  \end{array}\right).
\end{equation}
Unitarity and microreversibility\cite{ButtikerReciprocity} impose the following conditions on the scattering probabilities:
\begin{eqnarray}\label{eq:unitarity}
  &&T_{tc}+T_{bc}+R_s=1,\nonumber\\
  &&T_{ct}+R_{tt}+R_{bt}=1,\nonumber\\
  &&T_{cb}+R_{bb}+R_{tb}=1,\nonumber\\
  &&T_{cb}+T_{ct}+R_s=1,\nonumber\\
  &&T_{bc}+R_{bb}+R_{bt}=1,\nonumber\\
  &&T_{tc}+R_{tt}+R_{tb}=1.
\end{eqnarray}
Eqs.~(\ref{eq:unitarity}) demonstrate that there are only three independent scattering probabilities which
fully characterize a lead/coherent sample junction.  We choose these to be
$R_{tt}\equiv R_t,\,R_{bb}\equiv R_b$, and $ R_{tb}=R_{bt}\equiv R_A$.
As a convenience we will assume in our discussion that all contacts are identical;
differences between contacts could readily be accounted for in interpreting experiments if they
were adequately characterized.

The way a particular contact connects to the outer world specifies additional relations between `in' and `out' electrochemical potentials.
For example, when the top arm is open, one has $\mu_{t,in}=\mu_{t,out}$.  This applies equally well to voltage probes.
When a voltage probe is connected to the the top layer only, for example, we set $\mu_{t,in}=\mu_{t,out}$ as usual.
However we also set $\mu_{b,in}=\mu_{b,out}$.  Even though the bottom layer chemical potential is not {\em read} externally
it is still true that no current is flowing to the outside through this arm.  Contacts that are not layer selective are
electrically connected to the same external reservoir and therefore can be represented by setting $\mu_{t,in}=\mu_{b,in}$,
whether used as simply as voltage probes or as a current source or drain.
When current is allowed to flow between contact arms through an external resistor, the difference between their chemical
potentials is proportional to this current.
The unitarity and microreversibility relations, Eqs.~(\ref{eq:unitarity}) ensure that the voltage probe
readings made by probes that contact both layers or either layer individually agree.
\begin{figure}
\begin{center}
\includegraphics[scale=0.3,bb=0 0 562 563]{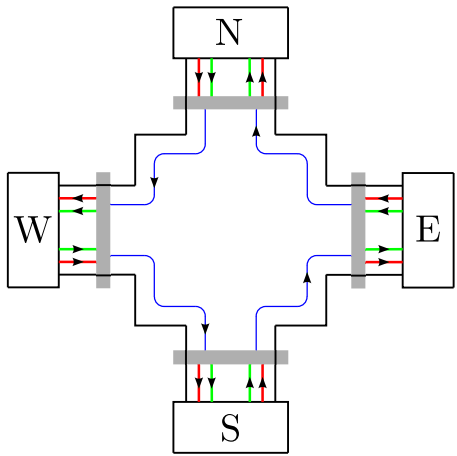}
\caption{(color online) Schematic of a four-probe measurement. The active
area of the bilayer is surrounded by a single current-carrying
edge-state channel shared between the two
layers (solid blue line). Each of the arms connecting the active area to the North, East, West, and South leads has a set of edge states described in Fig.~\ref{fig:edgestates}. There are a large variety of possible
measurements in which two of the four contacts are used as source and drain and two as
voltage probes.  Contacts are normally connected to only one of the two layers, as described in the text.}
\label{fig:setup}
\end{center}
\end{figure}

%\subsection{Two-probe conductances and minimum tunneling resistance}
Having described the scattering problem near the contacts, we can proceed to calculate various experimentally relevant quantities.
In general the effective single-particle scattering problem for the quasiparticles must be solved together with equations for the condensate.
(The quasiparticle mean-field Hamiltonian and hence the
transmission probabilities depend on the condensate configuration, which is in turn
influenced by the transport currents.)
At small voltages and currents, however, we can neglect this dependence, and treat the transmissions and reflections as constants
dependent on the equilibrium condensate configuration.

The typical experimental setup is depicted in Fig.~\ref{fig:setup}.
In the two-probe case the additional voltage probes are absent and it does not matter which pair of contacts we take as a source and a drain.
The three reflection probabilities that characterize the contacts, $R_t,\,R_b$, and $R_A$
can be determined by three independent measurements which take advantage of the freedom that the current
at both source and drain can flow through only one layer or through both.
We take these experiments to be  {\em parallel flow},  {\em drag}, and {\em tunneling} measurements.
In {\em parallel flow} the layers are shorted at both source and drain. In  {\em drag}, current
flows in from and out through the drive layer; no current flows through the drag layer arms at
either source or drain, or obviously at any voltage probes which might be present.
Finally, in the {\em tunneling} measurement current is injected into one layer, and extracted from the other one.
In most of this paper we assume that the condensate is time independent.  For tunneling experiments
this requires that currents be smaller than a critical tunneling
current.\cite{Ezawa1994, Tiemann2008NJoP,Tiemann2009CrCurr, SuPRB2010,eastham_critical_2010}

We find that the parallel, drag, and tunneling linear conductances are given respectively by
\begin{eqnarray}\label{eq:conductances}
  G_{\parallel}&=&\frac{1-R_s}{1+R_s}\frac{e^2}{h},\nonumber\\
  G_{drag}&=&\frac{1-R_t-R_b-R_A^2+R_b R_t}{1-R_t+R_b+R_A^2-R_b R_t}\frac{e^2}{h},\nonumber\\
  G_{tunnel}&=&\frac{1-R_t-R_b-R_A^2+R_b R_t}{1+R_A^2-R_b R_t}\frac{e^2}{h}.
\end{eqnarray}
In the above expression for $G_{\parallel}$ the skipping probability, $R_s$, is given by $R_s=R_t+R_b+2R_A-1$, as dictated by Eqs.~(\ref{eq:unitarity}). The drag conductance was evaluated for the case of a floating drag layer, {\em i.e.} an uncontacted drag layer. The same expression holds in the case of a grounded drag layer with a large (compared to the critical tunneling current) total current flowing through the bilayer. When the drag layer is grounded, the condensate phase will be time-dependent as discussed in the next Section.

One might view as ideal the situation in which the two-layers behave identically, the skipping probability is small, and there is no memory of incoming layer
when reflection by the active coherent region occurs.  These conditions together imply that $R_s=0$, $R_t=R_b=R_A=1/4$ and
give ideal values for $G_{\parallel}= e^2/h$, and for $G_{drag} = G_{tunnel}= 0.5 \, e^2/h$.
The parallel flow two-contact conductance is the usual ideal quantum Hall effect value, while the
drag and tunnel conductances are half as large.
The maximum possible tunnel conductance is actually somewhat larger
and reached at $R_t=R_b=0, R_A=1/2$, when it becomes $G^{max}_{tunnel}= 0.6 \,e^2/h$.
For these values $G^{drag} = G^{max}_{tunnel}$.
Vanishing normal reflection and maximum Andreev reflection seem unlikely in semiconductor bilayers,
since reflection in the same layer should be, if anything, more likely than reflection to the other layer.
A more plausible extreme set of values might be $R_s=0$, $R_t=1/2$ and $R_b=1/2$
which yields $G_{\parallel} = e^2/h$, and $G_{drag}= G_{tunnel}=  1/3 \, e^2/h$.
Experimental two-probe tunnel resistances in the neighorhood of $3 \frac{h}{e^2} \sim 100\,\textrm{kOhm}$
are indeed commonly observed in coherent bilayers.
Even for a strong static condensate the
maximum tunneling conductance is always associated
with a large resistance, which can be viewed as a contact resistance.
In this sense there is no ideal dissipationless Josephson effect in semiconductor bilayers.
The finite resistances do not necessarily, as it is often assumed,\cite{SternTunnel,Balents2001,Fogler2001,stern_strong_2002}
imply that the condensate is lost because of disorder and/or quantum fluctuations.
We will return to this point in the the discussion section.
Measurements of two-probe linear conductances for several different contacting geometries
can determine the numerical values of $R_{t}$, $R_{b}$ and $R_{A}$, and in this way
shed light on the nature of the lead/active region junction.

\section{Time-Dependent Condensates:  The Case of Grounded Drag\label{sec:timedepcond}}

So far we have been concerned with problems with a time-independent condensate. There is a class of experiments, however, in which
the contacting geometry and bias voltages lead to a net current flow between layers which exceeds the
critical value.  The interlayer current flow can either be forced by
the contacting geometry, as in counterflow and tunneling experiments, or implied by quasiparticle transport theory.
An example of the second case is provided by a drag experiment in which the drag layer is grounded, {\em i.e.} it is
connected to the drain contact rather than floating as in the classic drag experiment.
The description of the two types of experiments is slightly different: in counterflow and tunneling, there \emph{must} be interlayer current, and the full system of equations for the condensate and quasiparticles needs to be solved to find it, as well as the corresponding distribution of interlayer voltages. In the drag experiment with a grounded drag layer, on the other hand,
there is always a current path connecting source and drain that lies in the drive layer only.
However if a naive Landauer-Buttiker description with a time-independent condensate predicts an interlayer current
larger than the condensate can deliver, there must be an interlayer voltage to ensure that almost all the current flows through the drive layer. We leave calculating the full I-V curve for tunneling experiments to future work,\cite{DAPAHMinprep} and concentrate here on a simpler case of a drag experiment with the drag layer grounded at the drain. We will show that the measurement of the interlayer voltage generated
in this type of measurement can provide a test of the strong-coherence theory developed in the previous Section.

For definiteness, let as assume that electrons are injected into the West bottom layer
in Fig.~\ref{fig:setup}, and extracted from both North top and bottom layers, going counter-clockwise along the sample edge. Application of
the Landauer-Buttiker formalism predicts that the current that exits through the drag (top) layer, $I_{drag}$,  is
\begin{equation}
 I_{drag}=\frac{(1-R_A-R_t)}{2-2 R_A-R_t-R_b} \; I_{tot},
\end{equation}
where $I_{tot}$ is the total current through the system.  When this current exceeds the critical tunneling current
the condensate cannot be stationary\cite{su_to_2008} and a
finite interlayer voltage, developed through interlayer polarization, should be measurable.  This additional voltage
will reduce the current flowing to ground through the drag layer to a value below the critical current.
Experimentally, the total current is quite often much larger than the critical tunneling current.  In this case interlayer tunneling can be neglected altogether
and the interlayer voltage can then be found from the condition that the entire current flows through the drive layer. The latter condition follows from the fact\cite{DAPAHMinprep} that for a non stationary condensate collective tunneling becomes an ineffective sink for
condensate currents, and the interlayer resistance due to incoherent tunneling
is much larger than the resistance associated with the carrying the current in the drive layer only.

Since we neglect real interlayer electron transfer, we can remove the time dependence of the condensate by switching to a frame rotating with the order parameter. It must be noted that the distribution of interlayer electrostatic potentials has to be calculated taking full account of a bilayer electrostatics, modified by the presence of the condensate.\cite{moon_spontaneous_1995} However, for a well developed QHE the top and bottom edges running from West to North, and from North to West must be equipotentials. Moreover, the interlayer voltage must be equal at the above edges, otherwise there would be an ever-increasing counterflow from one edge to the other. In principle, this can be counteracted by counterflow decay due to real tunneling, which we take to be negligible at present, or phase slippage events,\cite{tinkham_introduction_2004} but these should not occur at small currents, since vortices are pinned. Hence we conclude that the rate at which the order parameter rotates, $\dot\varphi$, where $\varphi$ is its phase must be constant at the system edge, and it is this value of $-\hbar\dot\varphi=eV_b-eV_t$ at the edge that is measured by a voltage probe as an interlayer electrochemical potential difference. The latter statement assumes that the quasiparticle dynamics is fast on the condensate time scales, thus the quasiparticles are in equilibrium with the local instant value of the condensate.

Now let us calculate the value of the interlayer electrochemical potential difference that is necessary for the absence of the current through the drag (top) layer. Switching to a rotating frame corresponds to a gauge transformation $\Psi_{t,b}\to \exp(-ieV_{t,b}/\hbar) \, \Psi_{t,b}$, where $\Psi_{t,b}$ is the annihilation operator in the top and bottom layers, respectively.  This transformation removes the time dependence of the order parameter by shifting the electrochemical potentials in the top and bottom layers by $-eV_{t,b}$. Since the quasiparticles now see a time-independent spontaneous tunneling amplitude, we can apply Landauer-Buttiker formulae to the above shifted electrochemical potentials substituting $\mu_{in/out,t,b}\to \mu_{in/out,t,b}-eV_{t,b}$ in Eq.~(\ref{eq:chempots}). Note that $\mu_c$ remains unchanged. The electrochemical potential difference is found from the condition that there is no current flowing out of the top layer at North, and the \textit{unshifted} electrochemical potential in North top arm coincides with the drain electrochemical potential. The result is
\begin{equation}\label{eq:interlayerV}
  V_b-V_t=\frac{1-R_t-R_A}{1-R_t-R_b+R_t R_b -R_A^2}\frac{h}{e^2}I_{tot},
\end{equation}
where, as before, $I_{tot}$ is the current flowing through the bilayer. We note that the conductance in this case is still given by the corresponding expression in Eq.~(\ref{eq:conductances}). The interlayer voltage, Eq.~(\ref{eq:interlayerV}), depends on the same set of reflection probabilities that can be, in principle, extracted from the two-probe conductances measurements.  (See Eqs.~(\ref{eq:conductances})).
Confirmation of this prediction would provide strong confirmation that the strong coherence regime modeled in the previous section
can be achieved experimentally.

\section{Weak coherence: p-ology and longitudinal drag \label{sec:pology}}

When the charge gap responsible for the quantum Hall effect is reduced,
disorder enables quasiparticles to
deviate from perfect chiral routes around the sample perimeter and travel across the bulk.
Processes of this type, although always present at finite temperatures, are
extremely weak when the quantum Hall effect is well developed.
The bulk scattering events which lead to deviations from a perfect
quantum Hall effect are normally well characterized by
the value of the longitudinal resistance measured in a four-probe geometry.
In this section we show that in coherent bilayers, longitudinal voltages
can develop in both drag and drive layers when the current source and drain contacts are layer selective,
even when transport is perfectly localized at the edge and the quantum Hall effect is still perfect when the source
and drain contacts are not layer selective.  The longitudinal voltages emerge when
interlayer coherence is weak enough to permit evanescent conduction in the
{\em pair-breaking} bilayer edge channel, leading to a non-negligible layer-memory length.
Below we explain a model intended to describe electrical measurements
in this regime, which we refer to the weak coherence regime.

We start this discussion of deviations from the strong coherence case by briefly
reviewing the main assumptions made in the previous section.
There it was assumed that an electron arriving at a voltage probe travelled
in a single chiral channel localized along the edge of the active coherent region.
Under such an assumption, unitarity of scattering near the probe ensures that
selective voltage probes connected to either layer
measure the same electrochemical potentials.
On the other hand, electrons propagating along the aforementioned edge channel could have come from either top or bottom layers of, say, an
upstream current probe, and these upstream probes could have different electrochemical potentials in top and bottom layers.
In order to allow for the possibility of layer memory being retained between probes,
we have to weaken our assumptions about electron propagation between probes.
Losing \emph{which layer} memory is, of course, a process of quantum mechanical scattering, which happens on a certain length scale around a contact.
Up to now, we have been assuming that this length scale is small, which corresponds to what we are referring to as the
strong coherence regime.  As coherence is weakened, say, by increasing temperature or changing
the total filling fraction, one can imagine a situation when this
\emph{which layer} memory length becomes considerable and eventually comparable to the system size.
Indeed, as we explain below, there are good reasons to expect that disorder can cause this length
to become comparable to sample dimensions even when the normal quantum Hall effect is still reasonably well established.
Voltage probe readings will then depend on which layer is contacted.

Quasiparticle transport in this regime is most generally described by a phenomenology which allows all possible chiral transmission
coefficients between neighboring two-layer contacts.
The most general model consistent
with a perfect $\nu=1$ integer quantum Hall effect in the active region, when normally contacted, is one in which outgoing
electrons in a lead are either reflected within the same lead or transmitted to
the nearest downstream contact but satisfy the constraint
\begin{equation}
\sum_{ij} T_{ij}= 1.
\end{equation}
Below we will construct a one-parameter Landauer-Buttiker-type
parameterization to describe the weak coherence regime,
which is less general but physically transparent.
In order to capture layer memory we need to relax the
microscopic single-channel edge model for the active region edge. %where $l,l'$ label top and bottom layers.
In order to explain the main ideas we apply this approach mostly to the case of
drag experiments, although the same type of analysis could (and has been)
applied to tunnel, counterflow, and a wide variety of other contacting arrangements.

Longitudinal drag always accompanies weakening
coherence in bilayers, and has previously been analyzed using an effective medium
theory due to Stern and Halperin which associates longitudinal drag with loss of coherence in the bulk of the system.\cite{stern_strong_2002}
We show here that longitudinal drag should still be expected even when transport currents still flow perfectly chirally at the edge of the system
and the normal quantum Hall effect is still accurately quantized.
We expect our model to be most relevant at low temperature and small $d/\ell$
when the bulk effective medium theory appears to fail.\cite{tutuc_giant_2009}
Our approach provides a complementary understanding of the
large longitudinal drag resistances that can appear in coherent bilayers.

In what follows we assume that current is injected into and drained from the
bottom layer so that the top layer is the drag layer.  We assume that electrons flow counter-clockwise around the sample as in Fig.~\ref{fig:setup}. Quantities related to the edge running from the electron source to the electron drain (simply source and drain from now on) are labeled with the superscript ``+'', and those for the edge running from drain to source - with the superscript ``-''.
For the ``+''(``-'') edge, the coordinate along the edge, $x$, is counted from the source (drain). Correspondingly, the distances from source to drain (along ``+'' edge), and from drain to source (along ``-'' edge) are denoted $L_\pm$.
For simplicity, we take all contacts to be identical.
%We also disregard the (experimentally relevant) asymmetry between the electron density profiles in
%the two layers due to doping profiles and different distance to front and back gates.
Finally, we neglect the ``skipping'' probability $R_s$, and the ``Andreev'' reflection probability, $R_A$ (see Eqs.~(\ref{eq:unitarity})).
Complications which arise when these assumptions are relaxed will be discussed below.

To describe the loss of \emph{which layer} memory with a minimal set of parameters, we assume that electrons reflect in the same layer in the leads  with probability $R_{tt}=R_{bb}=1/2$. These values are appropriate (dictated by Eqs.~(\ref{eq:unitarity})) near the strong coherence limit if Andreev and skipping probabilities are neglected, and there is a symmetry between layers. To specify position-dependent transmissions, we assume that if a voltage probe is physically close to a lead, electrons retain full memory of the lead from which they originate, {\em i.e.} $T_{tt}(x=0)=T_{bb}(x=0)=1/2$, while $T_{tb}(x=0)=T_{bt}(x=0)=0$.  For a voltage probe located far from a lead, on the other hand,
the layer memory is completely lost, $T_{tt}(x=\infty)=T_{bb}(x=\infty)=T_{tb}(x=\infty)=T_{bt}(x=\infty)=1/4$.
The simplest  single-length-scale and layer-symmetric\cite{note_symmetry} \emph{ansatz} for the loss of \emph{which layer} memory is as follows:
\begin{eqnarray}\label{eq:transmissions}
  R_{tt}&=&R_{bb}=1/2,\nonumber\\
  T_{tt}(x)&=&T_{bb}(x)=\frac{1}{4}(1+e^{-x/\xi}),\nonumber\\
  T_{tb}(x)&=&T_{bt}(x)=\frac{1}{4}(1-e^{-x/\xi}),
\end{eqnarray}
where $\xi$ is the characteristic memory-loss length scale.
A similar \emph{ansatz} was used previously\cite{YoshiokaAHM} to describe bilayer transport at total
filling factor $\nu=2$.

Having defined the transmission probabilities, we can now write down the equations for the voltmeter readings.
Given top and bottom layer incoming electrochemical potentials $\mu^\pm_{t0}$ and $\mu^\pm_{b0}$, the equations for voltmeters readings along the edges are
\begin{eqnarray}\label{eq:chempot_xdep}
  \mu^\pm_t(x)=T_{tt}(x)\mu^\pm_{t0}+T_{tb}(x)\mu^+_{b0}+R_{tt}\mu^\pm_t(x), \nonumber\\
  \mu^\pm_b(x)=T_{bt}(x)\mu^\pm_{t0}+T_{bb}(x)\mu^\pm_{b0}+R_{bb}\mu^\pm_b(x).
\end{eqnarray}
When these equations are written in the form
\begin{equation}
  \left(\begin{array}{c}
    \mu^\pm_{t}(x)\\
    \mu^\pm_b(x)
  \end{array}\right)=
  U(x)\left(\begin{array}{c}
    \mu^\pm_{t0}\\
    \mu^\pm_{b0}
  \end{array}\right),
\end{equation}
the matrix $U$ satisfies the property $U(x)U(y)=U(x+y)$.
Thus, the presence of a voltmeter does not alter readings of any voltmeters placed downstream.
We can therefore characterize a given edge by a position dependent chemical potential.
(This property can be traced to our assumption in this section that Andreev scattering
in the contact regions is not important.)

To close Eqs.~(\ref{eq:chempot_xdep}) we need to relate $\mu^\pm_{t,b}(L_\pm)$ to $\mu^\pm_{t0,b0}$ and the current flowing through the system. Taking into account that $\mu^\pm_{t,b}(L_\pm)$ correspond to readings of voltmeters situated just before the West (source) and East (drain) leads, respectively, and that the top arm is open at East and West, we can apply Eqs.~(\ref{eq:chempot_xdep}) for $x=0$ to get the boundary conditions,
\begin{eqnarray}\label{eq:boundarycond}
  \mu^+_{t0}&=&\mu^-_t(L_-),\nonumber\\
  \mu^-_{t0}&=&\mu^+_t(L_+),\nonumber\\
  \mu^+_{b0}&=&\mu_{source},\nonumber\\
  \mu^-_{b0}&=&\mu_{drain}.
\end{eqnarray}
The expression for the current, defined as the one flowing through the bottom arm of West or East contacts is
\begin{equation}\label{eq:totalcurrent}
  I_{tot}=\frac{1}{2}\frac{e}{h}(\mu_{source}-\mu^-_b(L_-)) =\frac{1}{2}\frac{e}{h}(\mu^+_b(L_+)-\mu_{drain}).
\end{equation}

Eqs.~(\ref{eq:chempot_xdep}),~(\ref{eq:boundarycond}), and~(\ref{eq:totalcurrent}) allow for a full characterization of
electrical measurements on a bilayer under the conditions of weak coherence,
and imply a well defined $\nu_{tot}=1$ QHE when the layers are regularly contacted. The considerations above are simplified in several ways, but have the advantage that they leads to a theory with a single parameter, which can be fixed by one
measurement and compared with the wide variety of alternate measurements on the same sample that differ only by switching contacts between
top and bottom layers or using a probe to contact both layers.   More complete theories depend on aspects of the
contacts that are not characterized in typical experiments.

In what follows we present examples of quantities that can be calculated in the present simple framework for the typical experimental geometry of Fig.~\ref{fig:setup}. We then compare our predictions to the experimental results of Refs.~[\onlinecite{KelloggPRL2002}], [\onlinecite{Kellogg2003}], and~[\onlinecite{tutuc_giant_2009}], where longitudinal and Hall drags are measured as
a function of total filling factor\cite{KelloggPRL2002,tutuc_giant_2009} or effective layer distance.\cite{Kellogg2003}

The longitudinal
drag resistance is normally defined in terms of the difference between probe voltage values measured in the drag layer and along the same edge. The Hall drag is defined similarly, except that the voltage probes are placed on opposite edges. The geometry of Kellogg \textit{et al.}~\cite{KelloggPRL2002,Kellogg2003} corresponds to that described in Fig.~\ref{fig:setup}. The longitudinal drag resistance is measured with source and drain being at the West and North contacts, while voltage probes situated at the South and East contacts, Fig.~\ref{fig:setup}. If we denote the length of the square mesa edge as $L$, in this setup $L_+=3L$ and $L_-=L$. The longitudinal drag resistance is then
$R_{xx}=(\mu^+_t(2L)-\mu^+_t(L))/eI_{tot}$.  The Hall resistance is measured between the South and North contacts
with the source and drain being at the West and East, respectively. Thus in the Hall drag measurement $L_\pm =2L$, and the Hall drag resistance is given by $R_{xy}=(\mu^+_t(L)-\mu^-_t(L))/eI_{tot}$. It is natural to express the resistances in terms of the parameter $p=\exp(-L/\xi)$, the probability of
retaining layer memory while transiting one edge of the sample, hence {\em p-ology}.  By solving Eqs.~(\ref{eq:chempot_xdep}),~(\ref{eq:boundarycond}), and~(\ref{eq:totalcurrent}) one obtains the following expressions:
\begin{eqnarray}\label{eq:resistances}
  R_{xx}&=&\frac{p(1-p)}{1+p+p^2+p^3}\frac{h}{e^2},\nonumber\\
  R_{xy}&=&\frac{(1-p)^2}{1+p^2}\frac{h}{e^2}.
\end{eqnarray}
The values of these drag resistances do not depend on whether the drag layer is grounded or not (as in Section~\ref{sec:timedepcond} we assume a current much larger than critical tunneling one flowing through the bilayer). Longitudinal drag resistance also does not depend on a particular chirality of electron motion around the sample (assumed counter-clockwise in Fig.~\ref{fig:setup}), while the Hall drag resistance changes its sign with a change in the chirality.

The configuration used for the measurement of the longitudinal drag resistance described above is also convenient to measure the four-probe interlayer resistance. This resistance is defined in the same way as the longitudinal drag resistance, except the voltage at the East contact is measured in the drive layer: $R_{inter}=(\mu^+_b(2L)-\mu^+_t(L))/eI_{tot}$. This resistance does depend on whether the drag layer is grounded or not, and direction of electron motion around the sample. To proceed we combine the present considerations with those from Section~\ref{sec:timedepcond}. Then in the case of a counter-clockwise propagation (as shown in Fig.~\ref{fig:setup}) we obtain
\begin{eqnarray}\label{eq:resis_interlayer}
  R^{floating}_{inter}&=&\frac{p}{1+p^2}\frac{h}{e^2},\nonumber\\
  R^{grounded}_{inter}&=&\frac{2+3 p+3 p^2}{1+p+p^2+p^3}\frac{h}{e^2},
\end{eqnarray}
while for clockwise propagation (\textit{i.e.} the opposite to the one shown in Fig.~\ref{fig:setup}) we get
\begin{eqnarray}\label{eq:resis_interlayer}
  R^{floating}_{inter}&=&-\frac{p}{1+p^2}\frac{h}{e^2},\nonumber\\
  R^{grounded}_{inter}&=&\frac{2- p- p^2}{1+p+p^2+p^3}\frac{h}{e^2}.
\end{eqnarray}
Here the superscripts specify the drag configuration that the interlayer resistances pertain to. We observe that in the case of a floating drag layer the interlayer resistance is an odd function of the magnetic field, while in the grounded drag layer case it does not have a definite parity for $p\neq 0$. We note once again that the expression for the interlayer resistance for the grounded drag configuration assumes that the total current floating through the bilayer is much larger than the critical tunneling one, which typically holds in drag experiments.

Finally, we can also make predictions for two-probe quantities, in particular, for how the values of conductances described in Section~\ref{sec:LB} change with deviation from the strong coherence limit. The most interesting cases measure tunneling conductances.
We assume a tunneling geometry that is identical to that of the Hall drag measurements described above, except that source and drain contacts are attached to opposite layers.  This leads to
\begin{eqnarray}\label{eq:condpology}
  G_{tunnel}(p)=\frac{1-p^2}{3+p^2}\frac{e^2}{h}.
\end{eqnarray}
This value applies to currents small compared to the critical one.

We emphasize that all expressions in Eqs.~(\ref{eq:resistances}), (\ref{eq:resis_interlayer}), and~(\ref{eq:condpology}) are expected to be more reliable close to strong coherence, $p\ll 1$, where the Quantum Hall effect in the parallel channel is well-developed, and transport is edge-dominated. For instance, the formal limit $p\to 1$ in the expression for $R^{grounded}_{inter}$ is meaningless. It is derived under the assumption that the condensate must be dynamical in the grounded drag case when the total current is much larger then the critical tunneling current, in order to prevent current from floating through the drag layer, as would be predicted by naive close-to-strong-coherence Landauer-Buttiker theory. (See Section\ref{sec:timedepcond} for the discussion of this point.)

We observe that $R_{xx}$ in Eq.~(\ref{eq:resistances}) is a positive non-monotonic function of parameter $p$, vanishing both for $p\to0$ (strong coherence) and $p\to 1$ (weak coherence).
It reaches a maximum of $\approx 4\textrm{KOhm}$ at $p\approx 0.35$, or $\xi\approx 0.9L$. The positiveness of $R_{xx}$ as defined above implies that the sign of the voltage drop in the drag layer is opposite to the sign of the longitudinal voltage in the drive layer, as found experimentally.

Since $R_{xx}$ and $R_{xy}$ depend on a single parameter, the relationship between $R_{xx}$ and
$R_{xy}$ is universal in this simplified theory.  One quantity of interest characterizing the rate at
which longitudinal drags grow as the Hall drag deviates from perfect quantization:
\begin{equation}
\lim_{p \to 0} \;  \left|\frac{dR_{x}}{dR_{xy}}\right| = 0.5.
\end{equation}
In Refs.~[\onlinecite{KelloggPRL2002}] and~[\onlinecite{Kellogg2003}] a maximum drag resistivity of $\approx 2\textrm{KOhm}$ was observed at
the lowest temperatures ($\sim 30$mK) both by varying the total filling and by varying the effective layer distance, and the extracted slope of the parametric $R_{xx}(R_{xy})$ dependence was $\approx 0.2$. So we see the in the case of aforementioned experiments the present theory predicts a longitudinal drag about twice as large as observed.
The discrepancy is most likely due to the fact that in experimental systems deviations from perfect parallel transport Hall conductance
and loss of layer memory appear together.  (This point is discussed at greater length in the following section.)  When the ordinary
Hall conductance is not perfectly quantized some of the transport is able to travel from source to drain without passing the
longitudinal Hall voltage probes, reducing the measured voltage difference.
Although Eqs.(~\ref{eq:resistances}) are not in quantitative agreement with experiment,
they do clearly demonstrate that longitudinal drag {\em must} accompany
deviations from perfect Hall drag quantization.  Large longitudinal drags do not imply phase separation into coherent and incoherent regions, and, in our view, this macroscopic phase separation is unlikely to be present in samples studied experimentally.

We note that the strong coherence limit Eq.~(\ref{eq:condpology}) gives a value of tunneling conductance close to the one observed in experiment at low temperature and small $d/\ell$(e.g. in Refs.~[\onlinecite{SpielmanPRL2000}] and~[\onlinecite{Tiemann2009CrCurr}]). However, the conductance deteriorates much faster with decreasing coherence than predicted by Eq.~(\ref{eq:condpology}),when the value of $p$ is taken from Hall drag measurements. (This dependence is reported in Ref.~[\onlinecite{SpielmanThesis}].) This may indicate that the critical current has decreased to a value smaller than the smallest measurement current levels used in these experiments.

\begin{figure}
\begin{center}
\includegraphics[scale=0.4,bb=0 0 488 509]{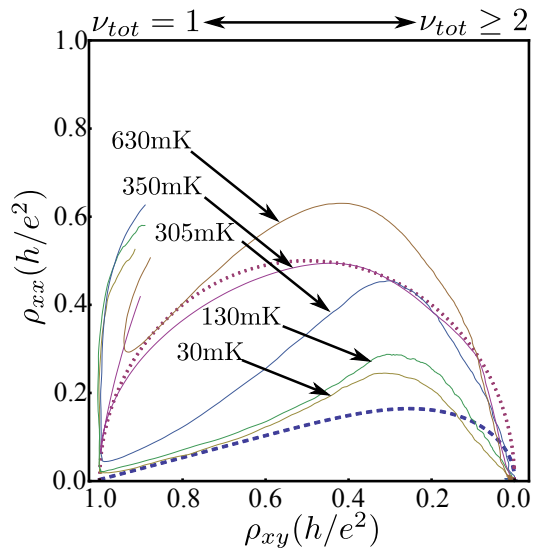}
\caption{(color online) Dependence of $\rho_{xx}$ on $\rho_{xy}$. Solid lines represent the experimental data of Ref.~[\onlinecite{tutuc_giant_2009}], where the filling fraction was varied from $\nu_{tot}\lesssim 1$ to $\nu_{tot}\gtrsim2$. The maximum in the longitudinal resistivity at low temperatures is reached around $\nu_{tot}\approx 1.15$ The dotted line corresponds to the semicircle law of Ref.~[\onlinecite{stern_strong_2002}], the dashed line is the prediction of this paper.}
\label{fig:resistances}
\end{center}
\end{figure}

In Ref.~[\onlinecite{tutuc_giant_2009}] a rectangular sample was used, and the longitudinal and drag resistivities per square,
$\rho_{xx}=\frac{W}{L}R_{xx}$ and $\rho_{xy}=R_{xy}$ ($W$ and $L$ being the width and length of the active area) were reported.
For the sample studied $W/L=1/4$. The explicit expressions for resistances for this case are not very instructive so
we present our findings graphically, comparing the parameter-free implicit dependence of
$\rho_{xx}$ on $\rho_{xy}$ predicted here, with the experimental results reported in Ref.~[\onlinecite{tutuc_giant_2009}]
and with the theoretical results in Ref.~[\onlinecite{stern_strong_2002}].
We note that close to the strong coherence limit ($|\rho_{xy}|\lesssim \frac{e^2}{h}$) and at low temperatures our curve appears to be close to the one observed in experiment, whereas at higher temperatures the one predicted by Stern and Halperin\cite{stern_strong_2002} is a
good match, as illustrated in Fig.~\ref{fig:resistances} (note a change in the sign convention for the Hall drag resistance). In the experiment of Ref.~[\onlinecite{tutuc_giant_2009}] the spontaneous coherence state is destroyed by changing filling factor, rather than effective layer separation. We still apply our contact model to predict the outcome of four-probe measurements though, as discussed in Section~\ref{sec:LB}.

We can conclude from this analysis that the presence of a non-zero longitudinal drag and an imperfectly quantized Hall drag alone do not
necessarily imply that inter-layer coherence is lost over a large fraction of the sample area.

\section{Landau-Zener Theory}
In this section we present a theoretical estimate of the value of layer-memory-loss length, $\xi$, introduced in the previous section, that
is based on a highly simplified model of electron propagation along the edge and intended to capture the essential roles of
layer asymmetric disorder and interlayer coherence. We map the problem of transport electrons propagating along an edge onto that of bilayer electrons in a single chiral charge-carrying channel which propagates at the magnetoplasmon velocity, $v_{emp}$.
Layer antisymmetric disorder and interlayer coherence are taken into account by fields that act on the
$z$ and $xy$ components of the {\em which layer} pseudospin degree-of-freedom.
The Schr\"{o}dinger equation for the problem reads
\begin{equation}\label{eq:1dSE}
  \left(-iv_{emp}\frac{d}{dx}+U_{dis}(x)\sigma_z +\Delta\sigma_x\right)\psi(x) =\varepsilon\psi(x),
\end{equation}
where, as before, $x$ is the coordinate along the edge. Without loss of generality, we can assume that the part of the exchange field due to the interlayer coherence has only a $\sigma_x$ component and for simplicity we take it to be spatially independent.
The $\sigma_z$ term in Eq.~(\ref{eq:1dSE}) accounts for the difference between the disorder potentials in the two layers; we assume that $U_{dis}$ changes sign on the disorder correlation length scale $\Lambda$, which is usually assumed to be comparable to the
distance to the dopant layer and therefore much longer than a magnetic length.

Because it is first order in derivatives, Eq.~(\ref{eq:1dSE}) maps to the time dependent Schr\"{o}dinger equation for a local spin in a time dependent magnetic field with components $(\Delta,0,U_{dis}(t=x/v_{emp}))$.
For weak coherence, the regime of present interest, $U\gg \Delta$, so that
the {\em local} eigenstates of Eq.~(\ref{eq:1dSE}) correspond to a definite layer index almost everywhere in space (
or at almost all times in the local spin mapping), except for branching points at which the disorder potential changes its sign.
Near such points, we can map Eq.~(\ref{eq:1dSE}) to the well-known Landau-Zener problem,\cite{Landau,Zener}
in which an electron has a probability to follow the instantaneous trajectory ({\em i.e.} retain its layer index) and
a probability of following the adiabatic path under the influence of $\Delta$,
changing its layer index.  The large value of $v_{emp}$ favors the former process can be dominant
and prevent an electron from losing its layer memory.

At each Landau-Zener branching point,  \emph{i.e.} each instantaneous level crossing,
the probability to follow the nonadiabatic path and stay in the same layer is
\begin{equation}
  g=e^{-\frac{2 \Delta^2 \Lambda}{\hbar v_{emp} U}}.
\end{equation}
Correspondingly, the probability to stay on the adiabatic path is $1-g$, which would mean retaining the layer index.
We introduce probabilities $p^{h,\ell}_n$ for an electron to be in the locally high and low energy states in a region where $|U_{dis}(x)|\gg \Delta$
after $n$ Landau-Zener branching events.  Using the branching rules we find that these probabilities satisfy
\begin{eqnarray}\label{eq:kineq}
  p^\ell_{n+1}&=&(1-g) \, p^\ell_n+g \, p^h_n,\nonumber\\
  p^h_{n+1}&=&(1-g) \, p^h_n+g \, p^\ell_n.
\end{eqnarray}
It follows that the transport quasiparticle layer polarization $P_n=(-1)^n \, (p^\ell_n-p^h_n)$
satisfies
\begin{equation}\label{eq:polarization}
\frac{dP}{dx}=-\frac{2(1-g)}{\Lambda} \; P(x)
\end{equation}
in the continuum limit.
We identify the layer-memory loss length with the characteristic length which
emerges from this analysis of polarization decay:
\begin{equation}
\label{eq:memorylength}
  \xi = \frac{\Lambda}{2(1-g)} \sim \frac{\hbar v_{emp} U}{\Delta^2}.
\end{equation}
Since the edge magnetoplasmon velocity in large Hall bar samples\cite{edgemagnetoplasmonref} has a
logarithmic enhancement related to the long-range of the Coulomb interaction
$\hbar v_{emp} / \ell \gg e^2/(\epsilon \ell)$.  Furthermore
$U$ can be $\sim e^2/\epsilon \ell$ due to weakened screening on Hall plateaus,
whereas $\Delta$ has a maximum value $\sim e^2/\epsilon \ell$ when coherence is complete.
From these estimates it is clear that the  layer memory length
on the right hand side of Eq.~(\ref{eq:memorylength}) must become
large as interlayer coherence is weakened, leading to longitudinal and Hall voltages whenever
current source or drain contacts are layer selective as explained in the previous section. These
layer memory effects can in principle
be separated experimentally from normal deviations from perfect quantum Hall effects, also
expected to become stronger when coherence weakens, by
comparing measurements made with layer selective and non-selective source and drain contacts.

\section{Discussion}

In this article we have developed a Landauer-Buttiker description of quasiparticle transport in
separately-contacted bilayer quantum Hall superfluids.  For well developed condensates we first assume that
current is carried between leads by a conducting channel consisting of quasiparticles that occupy both
layers simultaneously and have inter-layer phase coherence.  When this assumption applies,
voltages measured
along the sample edge have the same value for probes that contact both layers and probes that contact either of the
individual layers.  We refer to this regime as the {\em strong coherence} regime.  We show that two-probe conductances are then
dependent on the three independent Landauer-Buttiker scattering probabilities that characterize the link between two-layer leads and
the condensate edge.  Assuming that all contacts to the system are identical, these three parameters can be determined by
separate two-probe conductance measurements for parallel, drag, and tunnel contacting geometries.

The two-probe linear conductance values predicted in this approach tend to range from
$\sim 0.2\;e^2/h$ to $e^2/h$ depending on the contacting geometry and the contact scattering properties.
Indeed these values appear to be generally consistent with coherent bilayer
experiments performed at low temperature and small $d/\ell$ values.\cite{SpielmanPRL2000,SpielmanThesis,Tiemann2009CrCurr}
At higher temperatures and larger $d/\ell$ conductances tend to decrease.  A part of the decrease
is readily accounted for by deviations from a perfect quantum Hall effect due to scattering across the
bulk of the sample, which is always present at finite temperature and beyond the scope of the present theory.
Tunnel conductances decrease more rapidly than parallel transport or two-probe drive-layer conductances in drag experiments however. This observation points to the development of layer memory in
quasiparticle transport. We have therefore elaborated our strong coherence scattering formalism to allow
for layer memory, which we relate microscopically to disorder at the edge and resulting evanescent
transport in the pair-breaking bilayer edge-state channel.   This picture leads to a
layer memory length which is proportional to $\Delta^{-2}$, where $\Delta$ is the bulk
quasiparticle gap of the coherent state.  In addition to accounting for two-point tunnel geometry
conductances that are much smaller than $e^2/h$, this analysis shows that longitudinal
drag voltages and deviations from quantized Hall drag voltages develop when layer memory
lengths become comparable to distances between contacts.  We refer to this circumstance as the
{\em weak coherence} regime.  In this way we are able to
establish a relationship between Hall drag voltages and the longitudinal drag voltages which appear at weak coherence which is qualitatively consistent with experiment.\cite{KelloggPRL2002,Kellogg2003,tutuc_giant_2009} This picture also accounts for the sign of the longitudinal  drag voltage which is opposite to the sign of the longitudinal voltages
in the drive layer.

We have considered the transport properties of coherent bilayers in both time-independent condensate, and time-dependent condensate (interlayer current above the critical tunneling current) cases.  In either regime measured voltages reflect the properties of fermionic quasiparticle waves at the interface between the sample and bilayer leads. When the condensate is time-dependent, differences between chemical potentials measured by leads attached to different layers can reflect mainly this time dependence, and therefore the physics which controls its value in a particular geometry. Generally, the phase variation rate assumes a value sufficient to pump the required amount of current between layers via external electrical connections between layers, via inelastic quasiparticle tunneling between layers (the mechanism which appears to dominate experimentally at large bias voltage), or via local dynamic fluctuations of the condensate (as in the model used by Hyart and Rosenow,\cite{Hyart} discussed further below). In some cases a finite phase variation rate is necessary to make sure that current \emph{does not} flow between layers, as in the grounded drag configuration discussed in Section~\ref{sec:timedepcond} of this paper.

Before comparing the present considerations to related theoretical work, we would like to point out that a substantial part of the literature on the transport anomalies of coherent quantum Hall bilayers addresses tunneling\cite{SternTunnel,Balents2001,Fogler2001} and drag\cite{stern_strong_2002,SimonDrag, Roostaei} experiments separately.  In our view, an adequate understanding
of the superfluid transport anomalies can be achieved only by describing
all possible transport measurements within the same framework.  This point is emphasized further below.

One approach that is usually applied only to tunneling experiments accounts for collective fluctuations of bosonic particle-hole pairs.\cite{Balents2001, Fogler2001, SternTunnel, Hyart, Wen_ZeePRB} These theories assume that tunneling can be treated as a weak perturbative correction to a condensate with an interlayer phase precessing at an average rate proportional to the inter-layer potential $V$, {\em i.e.} $\varphi = eVt/\hbar$.  This picture is close to the one we employ when the tunneling current is much larger than its critical value, except that the fermionic quasiparticle degrees of freedom on which we focus do not appear explicitly. We return to this point below.  A perturbative treatment of tunneling can be valid only when the
system does not possess {\em spontaneous} long-range order.  Indeed the absence of spontaneous long range
order, even at $T=0$, appears likely in this system because of the presence of
meron condensate textures that are known to be induced by potential fluctuations.
A recent careful examination of a large body of experimental data by Hyart and Rosenow\cite{Hyart} has shown
that tunneling and closely related loop counterflow tunneling experiments can be accounted for
by a weak-tunneling bosonic picture with phenomenological phase correlation lengths and times,
presumed to be related respectively to the typical distance between merons and the time scale of their fluctuations.
This picture leads to a non-linear bilayer tunneling I(V) curve with a characteristic current scale $I_0$ proportional to
the bare tunneling amplitude $ \Delta_{SAS}^2$, a maximum differential conductance at zero bias that is proportional
to $I_0 \tau_{\phi}$, where $\tau_{\phi}$ is the phase fluctuation correlation time, and a region of
sharply negative differential resistance.  In comparison, the picture which underlies the present work is
one in which the characteristic current is a critical current below which the condensate is time-independent, and
the differential conductance at zero bias is limited by quasiparticle scattering properties to
values below around $0.1\;{\rm M \Omega}^{-1}$ in the strong coherence regime and to smaller values in the weak coherence regime.  We will argue below that the characteristic current scale should be the same size in both pictures. Nevertheless it should be possible to distinguish the two pictures experimentally.

As we have previously emphasized,\cite{su_to_2008,SuPRB2010} we have assumed that the condensate can be time-independent if the net current carried between layers inside the active region of the coherent bilayer is sufficiently small. When the condensate is time-independent, voltage measurements reflect only elastic quasiparticle scattering rates. For example all drag transport measurements (including drag-counterflow\cite{su_to_2008})
in which there is no net current flow between layers can be consistently understood in terms of the scattering properties of quasiparticles alone.
For tunneling, series counterflow,\cite{SuPRB2010} and other measurements in which there is net
current-flow between layers, the condensate can be static only if the supercurrent
injected into the sample by Andreev quasiparticle scattering can be sunk by time-independent coherent condensate
tunneling.  The critical current in essence\cite{SuPRB2010}  is the maximum possible value of
\begin{equation}
\label{eq:isink}
I_{sink} = \int_{A} d \vec{r} \; \frac{ e\Delta_{SAS} n}{2 \hbar} \sin(\phi)
\end{equation}
where $A$ is the full sample area.  Here $n$ is the condensate density
which is equal to $(2\pi\ell^2)^{-1}$ in the absence of disorder and quantum fluctuations.  It is quite clear that the experimental
critical currents, if non-zero, are much smaller than the value which would be obtained using this value of $n$ and
setting $\sin(\phi) \to 1$ everywhere.  It is natural to ascribe the large
reduction in critical currents to the presence of the merons,\cite{moon_spontaneous_1995}
which are known (see Refs.~[\onlinecite{eastham_critical_2010}] and~[\onlinecite{Roostaei}] and references therein) to be induced by disorder in coherent bilayers
and assumed to be present in all experimental samples.  Since the phase of the condensate winds by
$2 \pi$ around a meron core, setting the spatially average value of $\sin(\phi) \to 0$, it is clear that
the presence of merons will drastically influence the critical current. It seems clear that the
presence of disorder in bilayers implies that long-range spontaneous phase order is not present even at $T=0$.
As first emphasized in Ref.~[\onlinecite{SpielmanThesis}], and later detailed in Ref.~[\onlinecite{Hyart}], the evolution of transport anomalies with in-plane fields
suggests a correlation length $\xi_d$, possibly related to the distance between merons, which is comparable to the
spatial separation distance between the bilayer and the set-back dopant layer.  Long-range phase
order will, however, certainly be induced in response to the non-zero inter-layer tunneling amplitude
present in every sample.  Assuming that the degree of long-range order is determined by a balance
between the coupling of the condensate to inter-layer tunneling, and the exchange fields associated with meron-induced
condensate phase variation, we conclude that the effective condensate density is reduced from its
uniform system value by a factor of $\Delta_{SAS}n/(\rho_S \xi_d^{-2})$, where $\rho_S$ is the condensate stiffness.
With this modification $I_{sink}$ in Eq.~(\ref{eq:isink}) gives as a critical current, the same characteristic interlayer current value that emerges from the perturbative condensate tunneling theory.
We remark that this critical current estimate is not sensitive to the detailed meron-core configuration,
so thermal or quantum fluctuations in meron positions will produce small relative fluctuations in critical current values.

In summary, the transport anomalies of coherent quantum Hall bilayers
are similar to those of superconductors connected to normal metal leads, but are enriched by the possibility of contacting the two-layers separately, and by the presence of processes including transport through external conducting links, which violate separate particle-number conservation in the two layers.  This paper describes a conceptual framework for describing these anomalies which combines a Landauer-Buttiker picture of quasiparticle transport with a description of pseudospin-pumping by time-dependent condensates. Coherent bilayers are very sensitive to disorder because it tends to induce merons (vortices) in the order-parameter field and limit the maximum counterflow supercurrents which can be sunk by  coherent interlayer tunneling.  We have argued that critical currents remain finite even when merons are present because of the distortion of meron configurations by inter-layer tunneling.

\acknowledgements

The authors acknowledge helpful interactions with W. Dietsche, P. R. Eastham, J. P. Eisenstein,  A. D. K. Finck, B. I. Halperin, K. von Klitzing, B. Rozenow, L. Tiemann, E. Tutuc, and Y. Yoon. We are grateful to J.P. Eisenstein, A. D. K. Finck, and E. Tutuc for providing their experimental data. The \emph{p-ology} discussion in Section~\ref{sec:pology} elaborates on ideas developed in collaboration with J.P. Eisenstein. This work was supported by Welch Foundation grant TBF1473, and the NSF-NRI SWAN project.

%\bibliography{QHB}

\end{document}